# Polarization-Controlled Chiral Transport

Hang Zhu[1], Jian Wang[1], Andrea Alù[2,3] and Lin Chen[1,4,*]

[1] Wuhan National Laboratory for Optoelectronics and School of Optical and Electronic Information, Huazhong University of Science and Technology, Wuhan 430074, China

[2] Photonics Initiative, Advanced Science Research Center, City University of New York, New York, NY 10031, USA

[3] Physics Program, Graduate Center, City University of New York, New York, NY 10016, USA

[4] Shenzhen Huazhong University of Science and Technology Research Institute, Shenzhen 518063, China

* Corresponding author.

**Abstract**

Handedness-selective chiral transport is an intriguing phenomenon that not only holds significant importance for fundamental research, but it also carries application prospects in fields such as optical communications and sensing. Currently, on-chip chiral transport devices are static, unable to modulate the output modes based on the input modes. This limits both device functionality reconfiguration and information transmission capacity. Here, we propose to use the incident polarization diversity to control the Hamiltonian evolution path, achieving polarization-dependent chiral transport. By mapping the evolution path of TE and TM polarizations onto elaborately engineered double-coupled waveguides, we experimentally demonstrate that different polarizations yield controllable modal outputs. This work combines Multiple-Input Multiple-Output and polarization diversity concepts with chiral transport, and challenges the prevailing notion that the modal outputs are fixed to specific modes in chiral transport, thereby opening pathways for the development of on-chip reconfigurable and high-capacity handedness-selective devices.

**Main text**

Wavelength, mode and polarization multiplexing are recognized as the most effective solutions to address the need for multi-channel parallel communications in optical interconnects, optical computing and optical sensing. Polarization, as an intrinsic degree of freedom of light, is playing an increasingly significant role in multitasking information transmission. Developing polarization channel devices is beneficial for enhancing the control of the optical degrees of freedom[1,2], expanding information transmission capacity[3,4], improving transmission efficiency[5], and reducing channel crosstalk[6]. Successful implementations include polarization-controlled directional coupling[6], valley-locked beam splitters[7], circular polarization-dependent chiral ring resonators[2], and multifunctional spin lasers[8]. These polarization channel devices provide valuable methods for designing reconfigurable optical routers and optical computing processors, thereby contributing to the development of programmable optical communication and optical computing network technologies.

In a different context, exceptional points (EPs), consisting of singularities that arise in non-Hermitian systems where eigenvalues and eigenvectors coalesce, have been raising significant interest in various physical disciplines, including electronics and photonics[9-14]. The introduction and control of gain and loss distributions in photonic systems have enabled researchers to study the fundamental of EPs and the associated non-Hermitian phenomena in a wide range of optical systems, including microcavities[15-18], waveguides[19-32], gratings[33,34], and photonic crystals[35,36]. Many intriguing phenomena, such as enhanced sensing[37-40], unidirectional invisibility[41,42], single-mode lasers[16,43], and chiral dynamics[21-34], have emerged due to the unique topological features around EPs, which not only are of importance from the fundamental standpoint, but also have been leading to various cutting-edge photonic technologies. Out of the many associated phenomena, dynamically encircling an EP in non-Hermitian systems, has been under intense spotlight recently due to its chiral response, where the final system state depends on the handedness of EP encircling. Chiral transport achieved in these optical systems shows significant potential applications, such as quantum computing, asymmetric optical switches[23], polarization controllers[22,30], optical isolators[44], and more. However, chiral transport devices are static and each output port is locked to a specific mode regardless of the input, which is not conducive to functional reconfiguration and the improvement of transmission capacity.

In this article, we overcome the challenge that the output modes are fixed to specific modes independent of input modes in chiral transport by introducing polarization diversity in the EP-encircling dynamics. We show that polarization diversity can be used to control the evolution direction in the Hamiltonian parameter space, thereby achieving polarization-dependent state outputs. We present experimental results by mapping the evolution trajectories of TE and TM polarizations onto elaborately engineered double-coupled waveguides, showing different output modes corresponding to different polarizations at telecommunication wavelengths. Compared to previous research, our study challenges the inherent understanding that the output mode cannot be controlled by the input, but is uniquely locked to a single mode in EP-encircling chiral transport, providing opportunities for the development of on-chip reconfigurable and high-capacity chiral transmission devices.

The evolution path around EPs has been conventionally realized using coupled dielectric waveguides of rectangular cross-section. In such systems, the output port of the co-designed chiral transport devices is locked to a specific mode regardless of the input incidence[21,23,28,31,33]. Changing the width of the rectangular waveguide influences the effective refractive indices of both TE and TM polarizations at the same time. Therefore, the associated Hamiltonian parameters undergo a co-directional evolution path for both polarizations. Here, instead we implement anti-directional evolution paths for TE and TM polarizations, achieved by introducing L-shaped waveguide cross-sections in double-coupled waveguides. We start by analyzing TE and TM modes supported by L-shaped silicon waveguides, as shown in Figs. 1(a, b). It can be seen that TE and TM polarizations are distributed differently, where TE and TM polarizations are mainly distributed in the central and right regions of the waveguide, respectively. Although their effective refractive indices increase when either the top or bottom width increases, their sensitivity to top and bottom widths differs. The TE polarization is insensitive to changes in the top width, but sensitive to changes in the bottom width. Conversely, the TM polarization is sensitive to changes in the top width, but insensitive to changes in the bottom width. We are thus able to alter the top and bottom widths so that the effective refractive indices of TE and TM polarizations change in opposite directions. If an L-shaped waveguide and a rectangular waveguide are combined to form double-coupled waveguides, as shown in Fig. 1(c), we can optimize the geometry such that the detuning of TE and TM polarizations change in opposite directions, resulting in polarization-controlled chiral transport. The geometrical parameters of the double-coupled waveguides can be found in Supplementary Note 1.

The double-coupled waveguides can be rigorously described with the evolution equation as

$$i\partial/\partial z|\psi\rangle = H|\psi\rangle \tag{1}$$

where the eigenfunction is expressed as $|\psi(z)\rangle = [a_1(z), a_2(z)]^T$, $a_1(z)$ and $a_2(z)$ are the amplitudes of modes in each waveguide, and the eigenstates $[1,1]^T$ and $[1,-1]^T$ correspond to the symmetric and anti-symmetric modes, respectively. The Hamiltonian can be written as

$$H(z) = \begin{bmatrix} \beta(z)+i\gamma(z) & \kappa(z) \\ \kappa(z) & -\beta(z) \end{bmatrix}. \tag{2}$$

Here, $\beta(z)$, $\gamma(z)$, and $\kappa(z)$ represent the degree of detuning, loss rate and coupling strength of the system, respectively. The Hamiltonian parameters ($\beta$, $\gamma$, and $\kappa$), associated with the waveguide geometry, can be theoretically retrieved based on coupled mode theory[45] and the Beer-Lambert-Bouguer law[32]. The two eigenvalues are $E = i\gamma/2 \pm \sqrt{\kappa^2 + (\beta+i\gamma/2)^2}$ and the associated eigenvectors $|X\rangle = \left[\sqrt{1\pm M}, \pm\sqrt{1-(\pm M)}\right]^T/\sqrt{2}$, where $M = (\beta+i\gamma/2)/\sqrt{\kappa^2 + (\beta+i\gamma/2)^2}$, indicating that the system has an EP at $(\beta/\kappa, \gamma/\kappa) = (0, 2)$. Previous studies have shown that when the dynamic Hamiltonian trajectories surrounds the EP, the different losses experienced by different eigenstates lead to the generation of nonadiabatic transitions (NAT), which makes the output modes depend on the handedness of EP encircling. In Section I, the width of the right rectangular waveguide is kept constant, and the top and bottom widths of the left L-shaped waveguide reduces and increases along the $z$ direction. In this situation, the detuning of TE and TM polarizations between the two waveguides is negative and positive, respectively, as shown in Figs. 1(d-f). In Section II, the positions of the rectangular and L-shaped waveguides are swapped compared to Section I. Therefore, the detuning of TE and TM polarizations are opposite to those in Section I, respectively. To avoid path-dependent loss in encircling EPs, we have employed the strategy of "Hamiltonian hopping" between convergent eigenstates at Hamiltonian boundaries[28], i.e., points A$^-$, {B}, and A$^+$, as indicated in Fig. 1(c). A$^-$ (A$^+$) can be reached by increasing the gap distance, d, which effectively makes $\kappa(z) \to 0$. It should be noted that, TE and TM polarizations take opposite values at A$^-$ (A$^+$) points, due to their opposite detuning. A large loss rate $\gamma$ at {B} can be implemented using a semi-infinite slab waveguide to replace the first waveguide. Bend waveguides are used to connect the right waveguide with the slab waveguide. Once the guided waves reach the slab waveguide through the bend, light is not reflected back, i.e., $\gamma/\kappa \to \infty$ as required at {B}. As a result, the whole evolution trajectories are opposite for both polarizations, as shown in the right panel of Fig. 1(c). This presents a sharp contrast to previous studies, where the evolution trajectories were independent of the incidence[23-26,28,31-33]. More details regarding the dependence of $\gamma$ and $\kappa$ on the geometry of the double-coupled waveguides can be found in Supplementary Note 1.

To validate the functionality of polarization-controlled chiral transport, we conducted full-wave simulations using the finite-difference time-domain (FDTD) method to simulate the field intensity distributions of the double-coupled waveguides [Fig. 2]. The output mode is not locked to a specific mode, but it is dependent on the incident polarizations. For the left (right) port input with TE$_0$ and TM$_0$ modes, the output mode is TE$_1$ (TE$_0$) and TM$_0$ (TM$_1$) modes, respectively. When the TE$_0$ mode is input from the left port, most of the light energy resides in the left waveguide between A$^-$ and A$^+$, where the dominant eigenstate is $[0,1]^T$ [Fig. 2(a)]. However, when TM$_0$ mode is input, most of the light energy is concentrated in the right waveguide at A$^-$ and eventually dissipates as it travels through the bend waveguide to the slab waveguide, as shown in Fig. 2(b). When the $TE_0$ ($TM_0$) mode is input from the right port, the evolution process is similar to the one when $TM_0$ ($TE_0$) mode is input from the left port, and the output mode will be $TE_0$ ($TM_1$) mode [Figs. 2(c, d)]. It should be noted that the output mode is locked to $TE_1$ ($TM_0$) for the left port input and to $TE_0$ ($TM_1$) for the right port input, regardless of whether the input is a symmetric or antisymmetric mode. Simulation results with anti-symmetric mode incidence can be found in Supplementary Note 2. Overall, the simulation results demonstrate polarization-controlled chiral transmission for the double-coupled waveguides.

To understand the physics behind the intriguing polarization-controlled chiral response, the light transmission in the double-coupled waveguides can be rigorously described by Eq. (1). Assuming that $H(z)$ remains constant over the distance interval $[z_0, z]$, the final state can be written as

$$|\psi(z)\rangle = c_1(z_0)e^{iE_1(z-z_0)}X_1 + c_2(z_0)e^{iE_2(z-z_0)}X_2 \qquad (3)$$

Here, $E_1$, $E_2$ [Im($E_1$) ≤ Im($E_2$)], and $X_1$, $X_2$ are the eigenvalues and eigenvectors, respectively, extracted by solving Eq. 错误!未找到引用源。, $c_1$ and $c_2$ are arbitrary coefficients. The initial state is $|\psi(z_0)\rangle = c_1(z_0)X_1 + c_2(z_0)X_2$ at $z_0$. Equation 错误!未找到引用源。 indicates that the real and imaginary parts of the eigenvalues affect the phase and magnitude, respectively. We can thus present the dynamic Hamiltonian trajectories in the Riemann surfaces formed by the energy spectra of Hamiltonian for TE and TM polarizations, as schematically shown in Fig. 3. The dynamic trajectories are opposite for TE and TM polarizations, when they are injected into the same port, due to the opposite sign of the detuning β of TE and TM polarizations.

For symmetrical modes injected from the left port, the associated state $[1,1]^T$ at the starting point $(β/κ, γ/κ) = (0, 0)$ is situated in the upper half of the Riemann surface (Figs. 3a, b). When a $TE_0$ mode is injected and evolves clockwise from $(0, 0)$ to $A_{TE}^-$, $X_1$ is dominant, and it suffers from low loss as the imaginary part of $E_1$ is zero. The

Hamiltonian ultimately returns to (0, 0) after it undergoes successive hopping between $A_{TE}^-$, {B} and $A_{TE}^+$. Throughout the entire process, $X_1$ is always dominant, resulting in the output state $X_1 = [1,-1]^T$ on the lower half of the Riemann surface, corresponding to the $TE_1$ mode. In contrast to the $TE_0$ mode, the Hamiltonian evolves oppositely when the $TM_0$ mode is injected. The physical positions, associated with $A_{TE}^-$ and $A_{TE}^+$ for $TE_0$ mode, correspond to different values of the $TM_0$ mode, which we have denoted as $A_{TM}^-$ and $A_{TM}^+$, respectively. During the evolution process from (0, 0) to $A_{TM}^-$, the Hamiltonian evolves anticlockwise, in which $X_2$ is dominant and $X_1$ is slightly excited since the adiabatic condition cannot be strictly satisfied. During the hopping between $A_{TM}^-$, {B}, and $A_{TM}^+$, the dominant state $X_2$ incurs significant loss and eventually is dissipated, while $X_1$ remains lossless and becomes the dominant state, i.e., a NAT occur. The output state returns to $X_1 = [1,1]^T$ at (0, 0) on the upper half of the Riemann surface, corresponding to the $TM_0$ mode.

For symmetrical modes injected from the right port, the initial state $[1,1]^T$ is also situated in the upper half of the Riemann surface (Figs. 3c, d). For $TE_0$ injection, the dominant eigenstate transitions from $X_2$ to $X_1$, when the system experiences NAT. The system state becomes $[1,1]^T$ at the terminal point (0, 0), corresponding to $TE_0$ mode, as it evolves along the upper surface of the Riemann surface. For $TM_0$ injection, $X_1$ is consistently dominant and the system state finally output as $[1,-1]^T$ at (0, 0), corresponding to $TM_1$ mode. The dynamics process for the injection of symmetrical modes of different polarizations further validates the polarization-controlled chiral transport of the double-coupled waveguides. It should be noted that the output modes are locked with the same polarization incidence, regardless of the mode order. More details regarding the dynamics process can be found in Supplementary Note 3.

Scanning electron microscope (SEM) images for the double-coupled silicon waveguides in one of the fabricated samples are shown in Fig. 4a. The zoomed-in images on the left and right planes represent the regions bounded by the black rectangles (see Supplementary Notes 4 for the fabrication details and transmission measurement scheme). These images clearly indicate that the L-shaped and rectangular waveguides are reversed in the regions near the left and right ports. The simulated and measured transmission efficiencies for $TE_0$ and $TM_0$ input around 1550 nm are shown in Figs. 4(b-e). $T_{mn}$ ($T'_{mn}$) represents the transmission efficiency of $TE_m$ or $TM_m$ mode that outputs from right (left) port when $TE_n$ or $TM_n$ mode inputs from left (right) port. When $TE_0$ ($TM_0$) mode is injected from the left port, the extracted experimental efficiency deviation at 1550 nm is $T_{10} - T_{00} \approx 12$ dB ($T_{00} - T_{10} \approx 13$ dB), i.e., $T_{10} \gg T_{00}$ ($T_{00} \gg T_{10}$), indicating that $TE_1$ ($TM_0$) mode is dominate in the output. In contrast, when $TE_0$ ($TM_0$) mode is injected from the right port, we have $T'_{00} - T'_{10} \approx 28$ dB ($T'_{10} - T'_{00} \approx 22$ dB), i.e., $T'_{00} \gg T'_{10}$ ($T'_{10} \gg T'_{00}$), indicating that the output mode is dominated by $TE_0$ ($TM_1$) mode. These measurement results are consistent with the polarization-controlled chiral transport as have been predicted by the aforementioned theory and simulation. The experimental results generally match the simulation results, with some deviations mainly attributed to fabrication errors in the sample and testing inaccuracies due to noise, especially when the transmission efficiency is very low.

If further developed to guide different output polarizations along different waveguides, the polarization-controlled chiral converter can be extended to construct a chiral polarization router. This holds promise for applications in quantum walk systems to generate spin-correlated quantum states that are insensitive to input modes, indicating significant potential in quantum information processing[46]. Moreover, selectively introducing gain or loss in polarizations might be beneficial to achieve on-chip polarization-controlled output from lasers[47]. Compared to previous studies on chiral transport based on EP encircling[23-29], we have also introduced the degree of polarization freedom to increase the number of photonic transmission channels, which helps enhance multiplexing dimensions and expand communication capacity. Simulation results indicate that even with a ±50 nm etching overlay error, our device can maintain crosstalk below -10 dB, exhibiting favorable polarization-dependent asymmetric transmission effects. The device exhibits high fabrication robustness, facilitating its future practical fabrication and applications. The detailed simulation results are provided in Supplementary Note 5.

In conclusion, we have reported an on-chip polarization-controlled reconfigurable chiral transport device. We have shown that, in a double-coupled waveguide system, the Hamiltonian can be controlled to evolve around EPs in opposite directions for TE and TM polarizations, achieving polarization-controllable chiral transmission. Unlike previous works where output modes were uniquely determined by the incident direction, our approach allows output modes to be controlled by the incident polarization states. The silicon photonic experiments have verified the polarization-controlled chiral transmission effect. Our approach is general and can be extended to chiral transport in other physical fields such as acoustics, electronics, and condensed matter physics. From an application perspective, this novel device can function as a routing unit for high-performance optical communication and optical computing networks, effectively increasing optical information transmission capacity and enhancing optical computing power.

# Figures

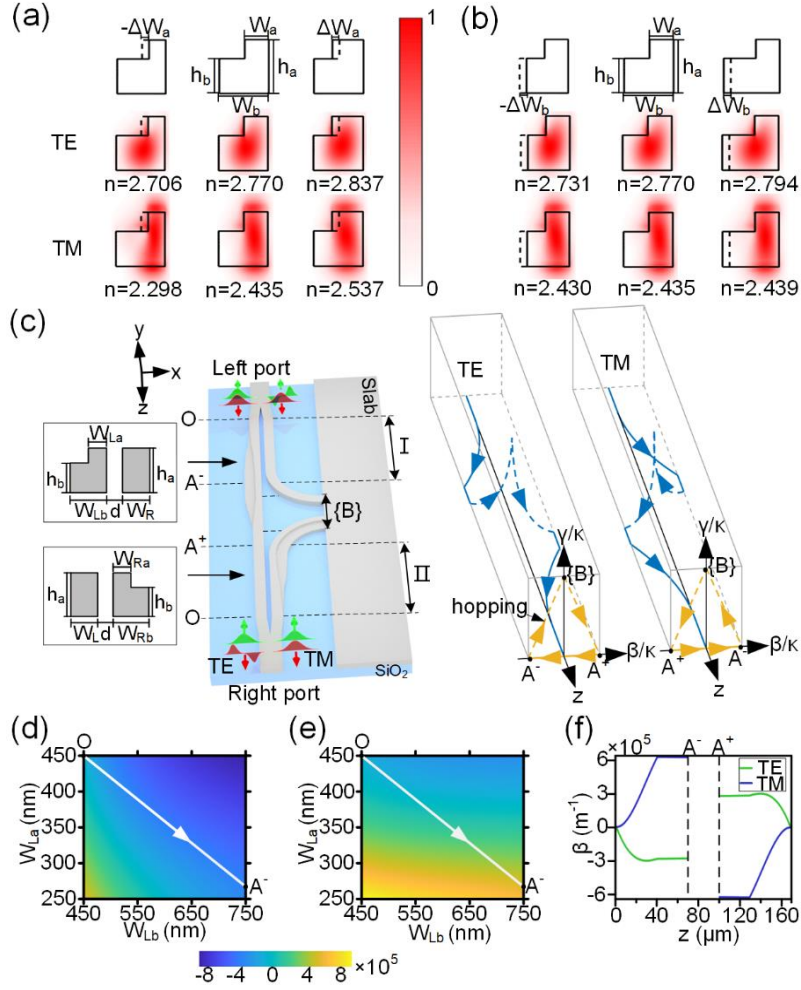

FIG. 1 Double-coupled waveguides for polarization-controlled chiral transport. (a, b) Modal field and index variations for TE and TM polarizations by changing the upper (a) and lower (b) widths of the L-shaped waveguide. Below the modal field distributions, *n* denotes the modal index. The lower and upper widths of the baseline waveguide are $W_b$ = 700 nm and $W_a$= 350 nm, respectively, and their width variations are $\Delta W_a = \Delta W_b$ =100 nm. (c) Dual-coupled waveguides for demonstrating polarization-controlled chiral transport. The cross-sectional parameters are marked in the left panel. The blue line with arrows in the right panel show the evolution trajectory, where its projection onto ($\beta/\kappa$, $\gamma/\kappa$) plane is marked by the brown line. (d, e) $\beta$ versus $W_{La}$ and $W_{Lb}$ at 1550 nm for TE (d) and TM (e) polarizations, with $W_R$ being fixed at 450 nm. The white straight lines represent the structural parameter variations from O to $A^-$. (f) $\beta$ as a function of the propagation distance, z. The entire silicon waveguides are covered by a 1-μm-thick $SiO_2$ layer, with the refractive indices of Si

and SiO$_2$ being of 3.478 and 1.444 at 1550 nm, respectively. The full height of L-shaped waveguide is h$_a$ = 340 nm, and the height of the lower waveguide is h$_b$ = 220 nm.

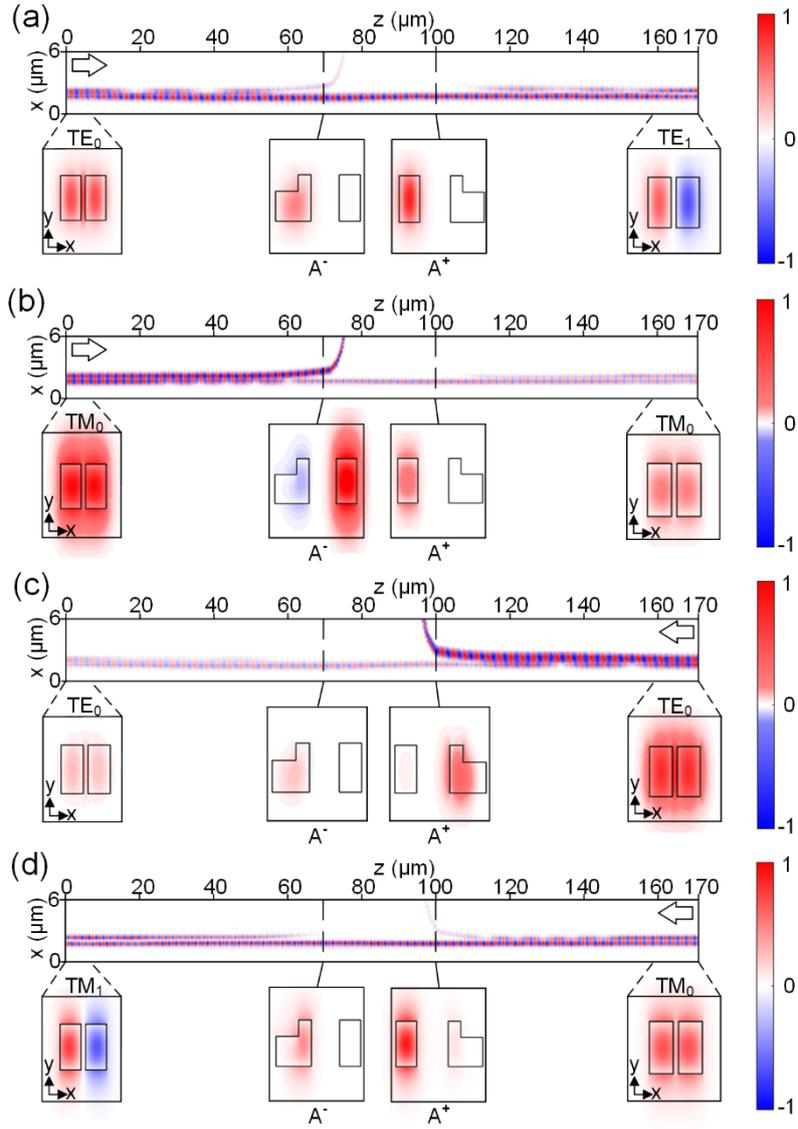

FIG. 2 Simulated field distributions. (a, c) Field distributions of $E_x$, when $TE_0$ mode inputs from the left (a) and right (c) ports, respectively. (b, d) Field distributions of $E_y$, when $TM_0$ mode inputs from the left (a) and right (c) ports, respectively.

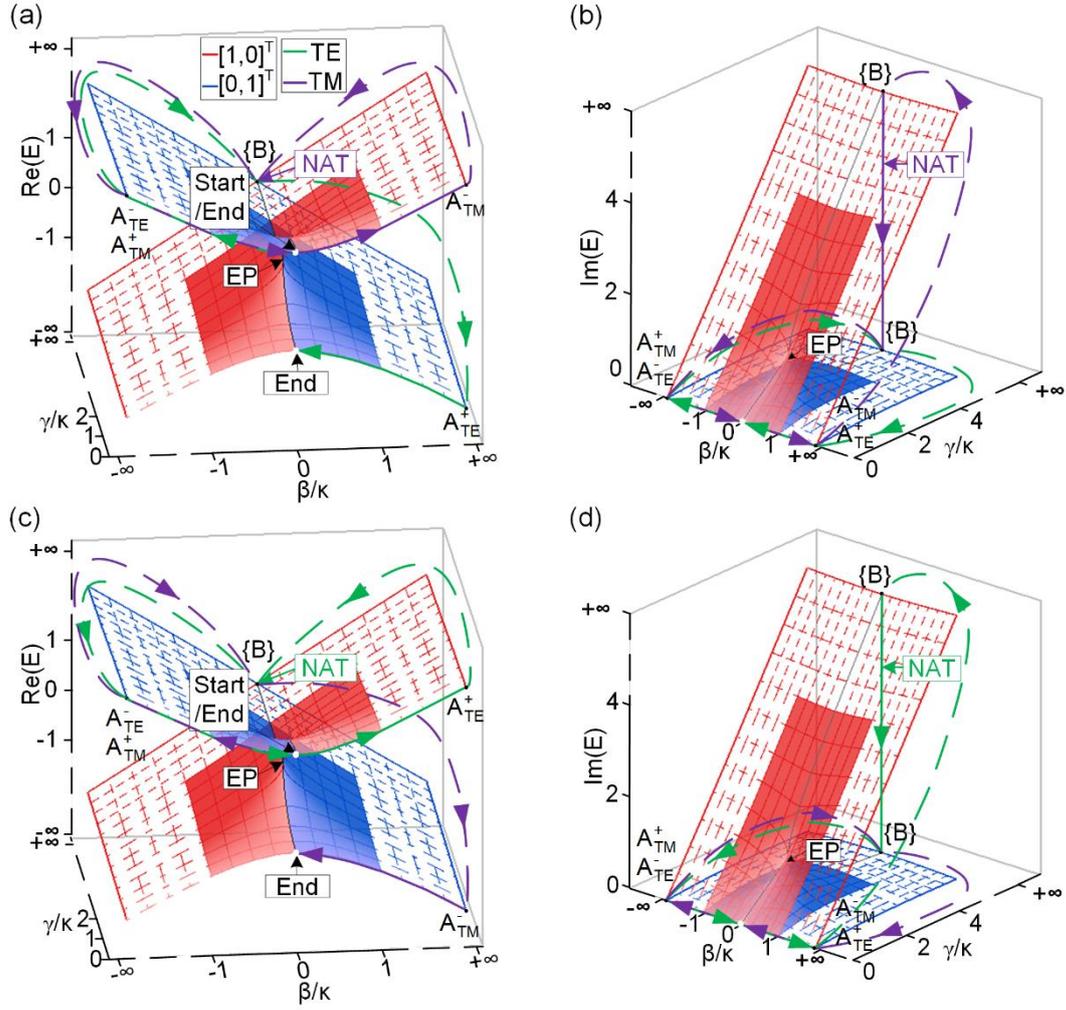

FIG. 3 System states evolving on the Riemann surfaces. The evolution trajectories in the Riemann surfaces formed by the real part Re(*E*) and imaginary part Im(*E*) of the energy spectra of *H*, when the symmetrical mode is input from the left (a, b) and right (c, d) ports, respectively. The dashed meshes represent the Hamiltonian parameters extended to infinity. The blue and red solid lines represent the parameter space boundary, associated with the eigenstates as $[0,1]^T$ and $[1,0]^T$, respectively. The dashed lines refer to Hamiltonian hopping among $A_{TE}^-$, $A_{TE}^+$, {B}, and $A_{TM}^-$, $A_{TM}^+$, {B}.

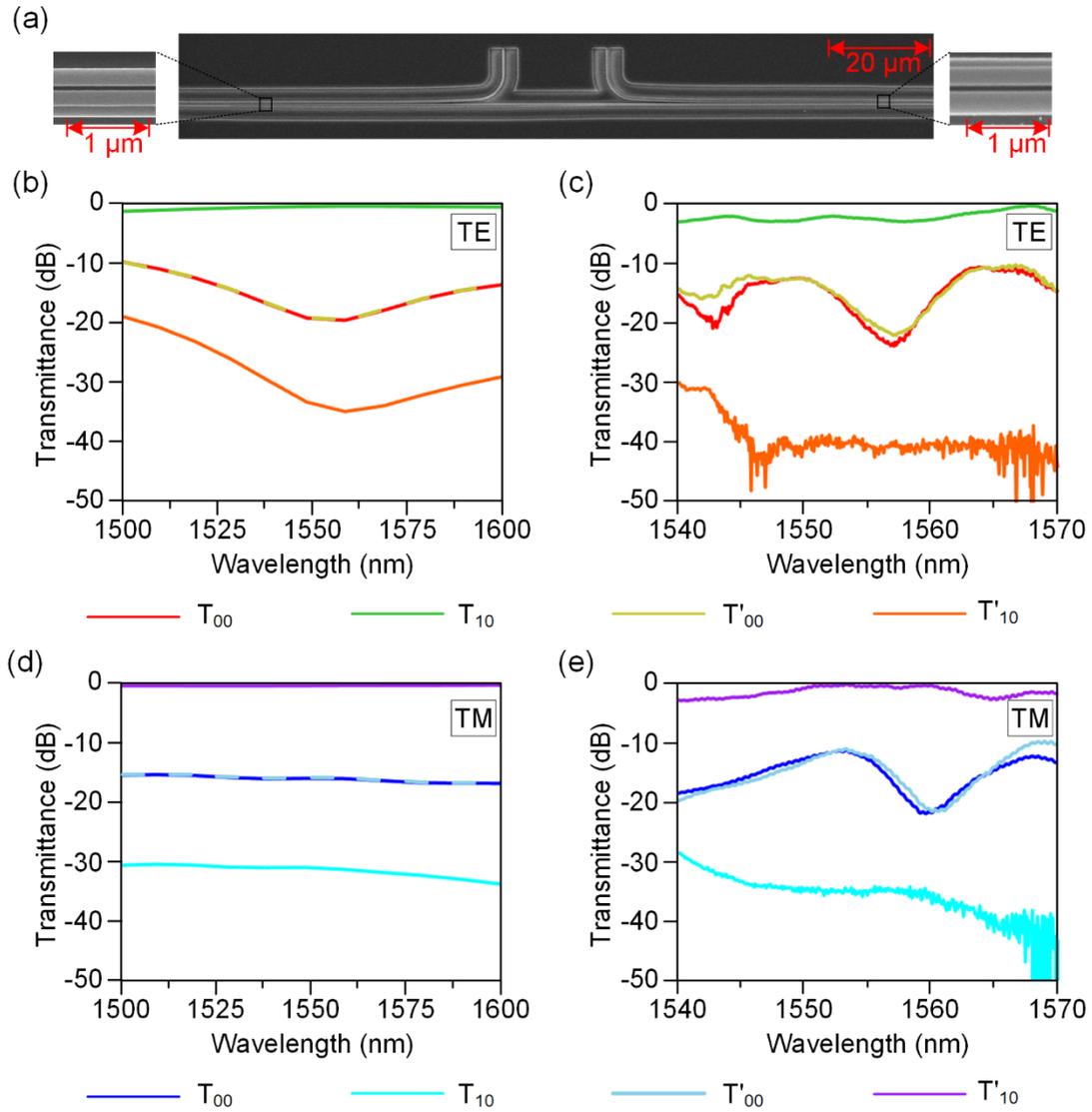

FIG. 4 Experimental demonstration. (a) SEM image of the device. (b, d) Simulated transmittance spectra at the output port over the wavelength range of 1500–1600 nm with $TE_0$ (b) and $TM_0$ (d) injection. (c, e) Experimental transmittance spectra at the output port over the wavelength range of 1540–1570 nm with $TE_0$ (c) and $TM_0$ (e) injection.

This work is supported by National Natural Science Foundation of China (Grant No. 12074137), National Key Research and Development Project of China (Grant No. 2021YFB2801903), and Science, Technology and Innovation Commission of Shenzhen Municipality (Grant No. JCYJ20220530161010023). A.A. was supported by the Simons Foundation. We thank the Center of Optoelectronic Micro&nano Fabrication and Characterizing Facility of WNLO for the support in SEM measurement.

**Supplementary Information for**

**Polarization-Controlled Chiral Transport**

**Supplementary Note 1: Structural and Hamiltonian parameters**

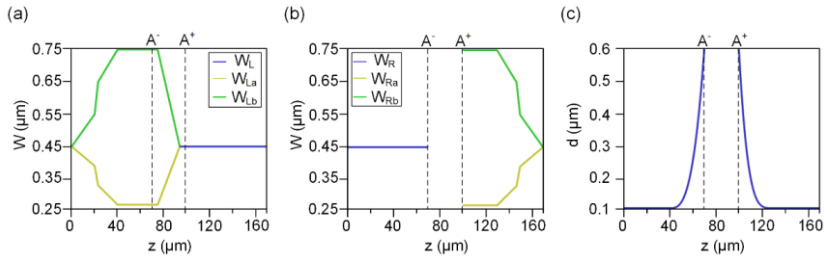

Supplementary Figure S1. Structural parameters of the double-coupled waveguides. The left (a) and right (b) waveguide widths, and gap distance (c) versus z.

The dependence of $W_L$ ($W_{La}/W_{Lb}$), $W_R$ ($W_{Ra}/W_{Rb}$) and d on z is depicted in Supplementary Figs. S1a-c, respectively. The corresponding variation of Hamiltonian parameters versus z is shown in Supplemental Fig. S2.

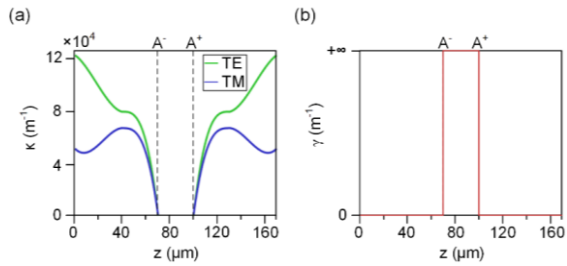

Supplementary Figure S2. The corresponding Hamiltonian parameters variation. (a) $\kappa$ and (b) $\gamma$ as a function of the propagation distance, z.

**Supplementary Note 2: Simulated results with anti-symmetric mode incidence**

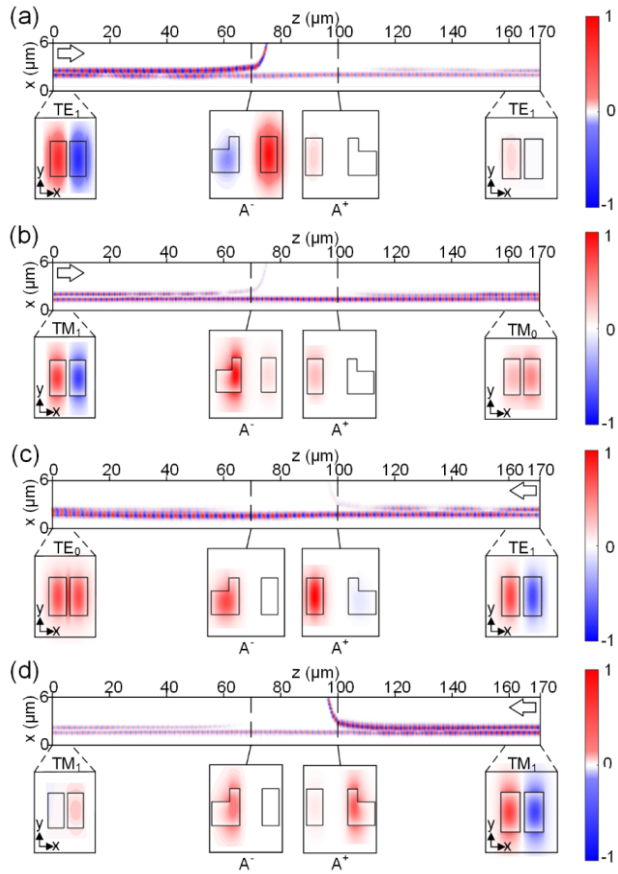

Supplementary Figure S3. Simulated field distributions. (a, c) Field distributions of $E_x$, when $TE_1$ mode inputs from the left (a) and right (c) ports, respectively. (b, d) Field distributions of $E_y$, when $TM_1$ mode inputs from the left (a) and right (c) ports, respectively.

**Supplementary Note 3: Dynamic process**

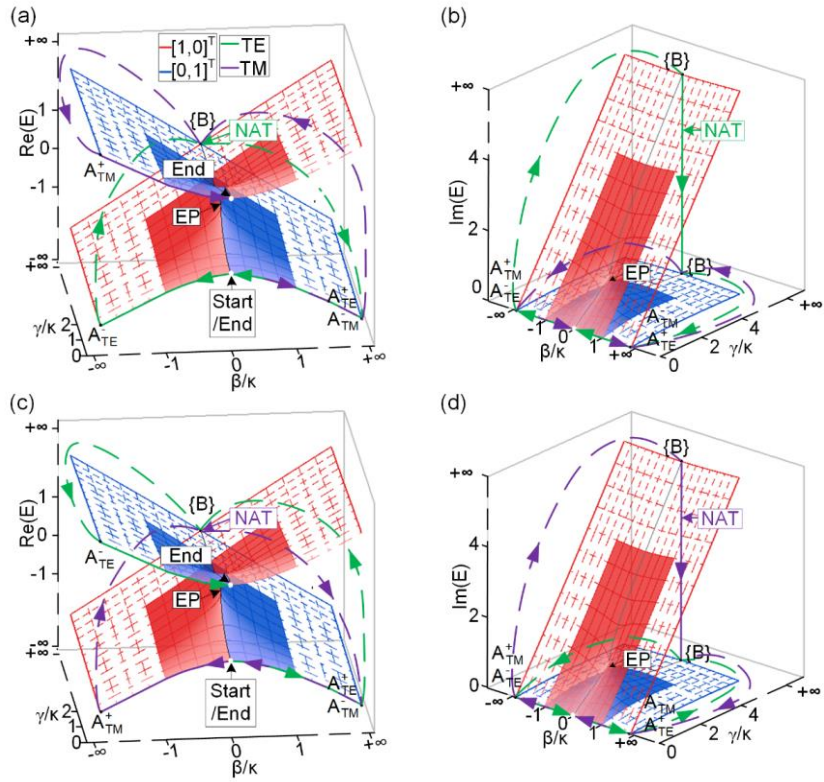

Supplementary Figure S4. System states evolving on the Riemann surfaces. The evolution trajectories in the Riemann surfaces formed by the real part Re(*E*) and imaginary part Im(*E*) of the energy spectra of *H*, when the anti-symmetrical mode is input from the left (a, b) and right (c, d) ports, respectively.

For anti-symmetrical modes injected from the left port, the initial state $[1,-1]^T$ at the starting point $(\beta/\kappa,\gamma/\kappa)=(0,0)$ is situated in the lower half of the Riemann surface (Supplementary Figs. S4a, b). When $TE_1$ mode is injected and evolves clockwise from $(0,0)$ to $A_{TE}^-$, $X_2$ is dominant and the other state $X_1$ is slightly excited. Because the dominant state $X_2$ incurs significant loss during "Hamiltonian hopping" from

$A_{TE}^-$ to {B}, it dissipates entirely. Meanwhile, $X_1$ remains lossless and becomes dominant from {B} to $A_{TE}^+$, i.e., NAT occurs. The output state returns to $X_1 = [1,-1]^T$ at $(0,0)$ in the lower half of the Riemann surface, corresponding to TE$_1$ mode. In contrast to TE$_1$ mode, the Hamiltonian evolves oppositely when TM$_1$ mode is injected. During the evolution process from $(0,0)$ to $A_{TM}^-$, $X_1$ is dominant since it suffers from low loss as the imaginary part of $E_1$ is zero. After the Hamiltonian successively experiences hopping between $A_{TM}^-$, {B}, and $A_{TM}^+$, it ultimately returns to $(0,0)$. Throughout the entire process, $X_1$ is consistently dominant, resulting in an output state of $X_1 = [1,1]^T$ on the upper half of the Riemann surface, corresponding to TM$_0$ mode.

For anti-symmetrical modes injected from the right port, the initial state $[1,-1]^T$ is also situated in the upper half of the Riemann surface (Supplementary Figs. S4c, d). For TE$_1$ injection, $X_1$ is consistently dominant, and eventually evolves along the upper surface of the Riemann surface to become $[1,1]^T$ at $(0,0)$, corresponding to TE$_0$ mode. For TM$_1$ injection, the dominant eigenstate transitions from $X_2$ to $X_1$, due to the occurrence of NAT, and eventually evolves along the lower surface of the Riemann surface to become $[1,-1]^T$ at $(0,0)$, corresponding to TM$_1$ mode.

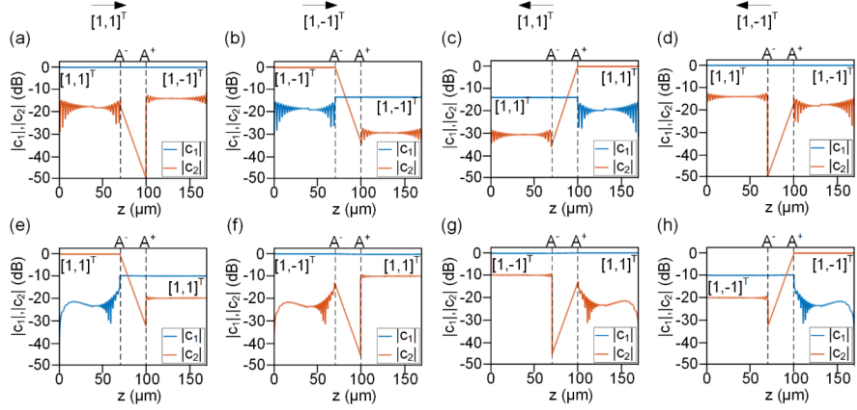

Supplementary Figure S5. Dynamics of the evolution trajectories for TE and TM polarizations. $|c_1|$ and $|c_2|$ represent the amplitude of the two eigenstates, $|X_1\rangle$ and $|X_2\rangle$. Coefficients $|c_1|$ and $|c_2|$ versus the propagation distance, z, for (a-d) TE and (e-h) TM polarizations. The left-pointing and right-pointing arrows indicate the injection of left and right ports, respectively.

In order to validate the chiral dynamics described in Fig. 3 in the main text, the required Hamiltonian parameters (Fig. 1f and Supplementary Fig. S2) are chosen to enable the evolution trajectories. Supplementary Fig. S5 illustrates the evolution of TE (a-d) and TM (e-h) modes, respectively. For TE polarization, the initial state is $[1,1]^T$ when it inputs from the left port. The dominant eigenstate is always the eigenstate, $|X_1\rangle$, and the final state is $[1,-1]^T$ (Supplementary Fig. S5a). When $[1,1]^T$ inputs from the right port, the dominant eigenstate is $|X_2\rangle$ for $z > A^+$, and shifts to $|X_1\rangle$ in the distance interval between $A^+$ and $A^-$ due to the emergence of NAT, and the final state returns to $[1,1]^T$ (Supplementary Fig. S5c). When $[1,-1]^T$ inputs from the left port, the finial state is $[1,-1]^T$ due to the emergence of NAT (Supplementary Fig. S5b). When $[1,-1]^T$ inputs from the right port, the finial stat is $[1,1]^T$ (Supplementary Fig. S5d).

For TM polarization, the initial state is $[1,1]^T$ when it inputs from the left port. The dominant eigenstate is $|X_1\rangle$ for z > A⁻, and shifts to $|X_2\rangle$ in the distance interval between A⁻ and A⁺ due to the emergence of NAT, and the final state returns to $[1,1]^T$ (Supplementary Fig. S5e). When $[1,1]^T$ inputs from the right port, the dominant eigenstate is always the eigenstate, $|X_2\rangle$, and the final state is $[1,-1]^T$ (Supplementary Fig. S5g). When the initial state $[1,-1]^T$ inputs from the left port, the finial state is $[1,1]^T$ (Supplementary Fig. S5f). When it inputs from the right port, the finial stat is $[1,-1]^T$ due to the emergence of NAT (Supplementary Fig. S5h).

**Supplementary Note 4: Fabrication details and measurement scheme**

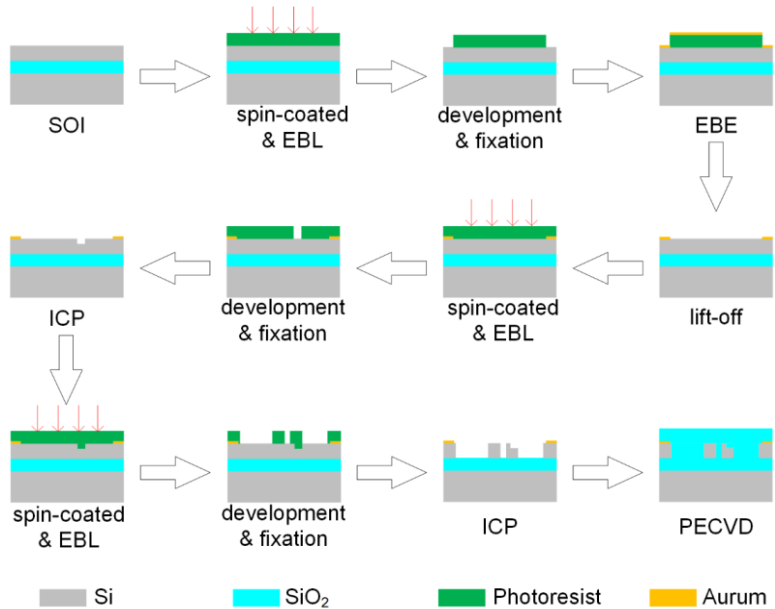

Supplementary Figure S6. Fabrication process of the samples.

Supplementary Fig. S6 shows the fabrication process of our device samples with a combination of three-step electron-beam lithography (EBL), inductively coupled plasma (ICP) etching, electron-beam evaporation (EBE), and plasma-enhanced chemical vapor deposition (PECVD).

Firstly, an SOI wafer was successively cleaned in ultrasound bath in acetone, isopropyl alcohol and deionized water, and then was dried under nitrogen flow. A 20-nm-thick Aurum layer with a 5-nm-thick Chromium adhere layer, was fabricated as the alignment marks by a first-step EBL, EBE and lift-off process. Photoresist was spin-coated onto the wafer surface and was patterned by EBL, which was followed by development and fixation. The Aurum marks were successively deposited by EBE, and the final alignment marks were formed by lift-off process. Secondly, the partially-etched layer of the L-shaped silicon waveguides was fabricated by using a second-step EBL and ICP etching. The photoresist was patterned by use of EBL, followed by ICP etching in which the etching time is precisely controlled to define the etching depth of

120 nm for the partially-etched layer. Thirdly, the fully-etched layer of the L-shaped silicon waveguides was fabricated by using a third-step EBL and ICP etching. The ICP etching time was controlled to define the 340-nm-height fully-etched layer after the photoresist was patterned by EBL. Finally, a 1-μm-thick $SiO_2$ layer is deposited by PECVD, to cover the entire sample for the optical field symmetry and structural protection.

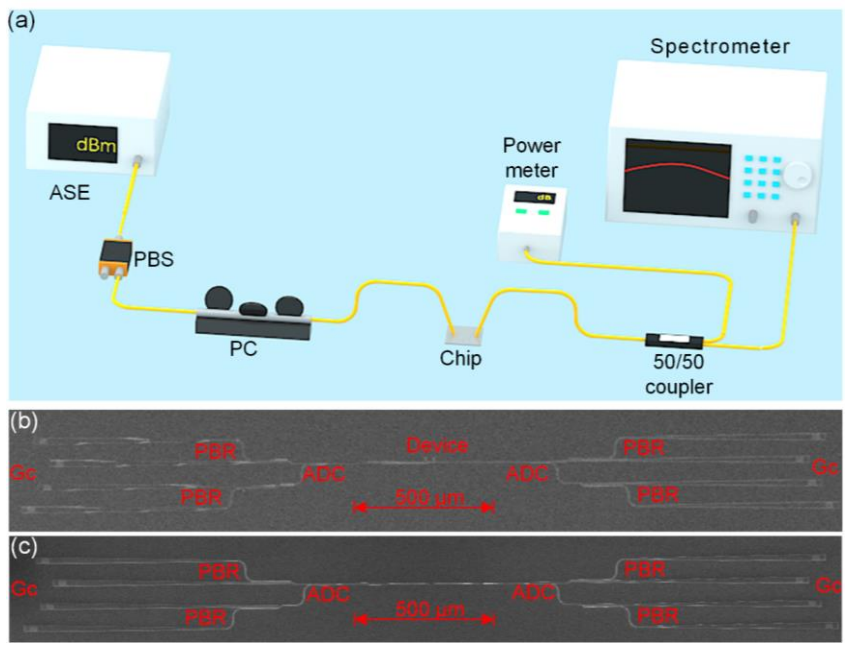

Supplementary Figure S7. Experimental demonstration. (a) The experimental configuration. (b) The SEM image of the fabricated sample consisting of the double-coupled silicon waveguides, GCs, ADCs and PBRs. (c) The SEM image of the control device without the double-coupled silicon waveguides.

Supplementary Fig. S7 presents the experimental setup for measuring the transmittance of the fabricated device. The near infrared light is provided by an amplified spontaneous emission (ASE) broadband light source (spectral range 1525 to 1600 nm). The polarization of the incident light is adjusted by polarization beam splitter (PBS) and polarization controller (PC) before light is coupled into the grating coupler (GC) through the fiber. The emergent light from the SOI chip is measured by the optical power meter and spectrometer. The optical power meter is used to adjust the angle between the fiber and the GC so as to maximize the coupling efficiency between them.

$TE_1$ ($TM_1$) mode separates from the straight bus waveguide and is converted into $TE_0$ ($TM_0$) mode by asymmetrical directional coupler (ADC). The polarization beam splitter and rotator (PBR) is used to generate or split $TE_0$ and $TM_0$ modes at the input and output ports. Supplementary Fig. S8 presents the simulated results of the PBR. Supplementary Fig. S8(a) shows the intensity distributions of the electric field when $TE_0$ mode inputs from port 1. The electric field intensity is close to 0 at port 3, indicating that only $TE_0$ mode outputs at port 2. Supplementary Fig. S8(b) shows the intensity distributions of the electric field when $TM_0$ mode inputs from port 1. The electric field intensity is close to 0 at port 2, and $TM_0$ mode is converted into $TE_0$ mode that outputs at port 3. Subsequently, the $TE_0$ modes from ports 2 and 3 are output to the detectors through the GCs. On the contrary, when light is coupled into the port 2 (port 3) with $TE_0$ mode through the GCs, it transforms into the $TE_0$ ($TM_0$) mode that outputs from port 1. Supplementary Figs. S8(c) and (d) show the transmittance of the $TE_0$ mode at port 2 and port 3, when the $TE_0$ and $TM_0$ modes input from port 1, respectively. The extinction ratio is defined as the ratio of the energy at the desired port to the energy at the undesired port. The estimated extinction ratios are larger than 35 dB over 1500−1600 nm, no matter when the $TE_0$ and $TM_0$ mode is injected, indicating that the PBRs hardly affect the measured transmittance of the double-coupled silicon waveguides.

The control device without the double-coupled silicon waveguides is used to evaluate the loss arising from ADCs, GCs and PBRs. The loss differences in Supplementary Figs. S7b, c can be used to extract the loss for different TE (TM) modes in the double-coupled silicon waveguides.

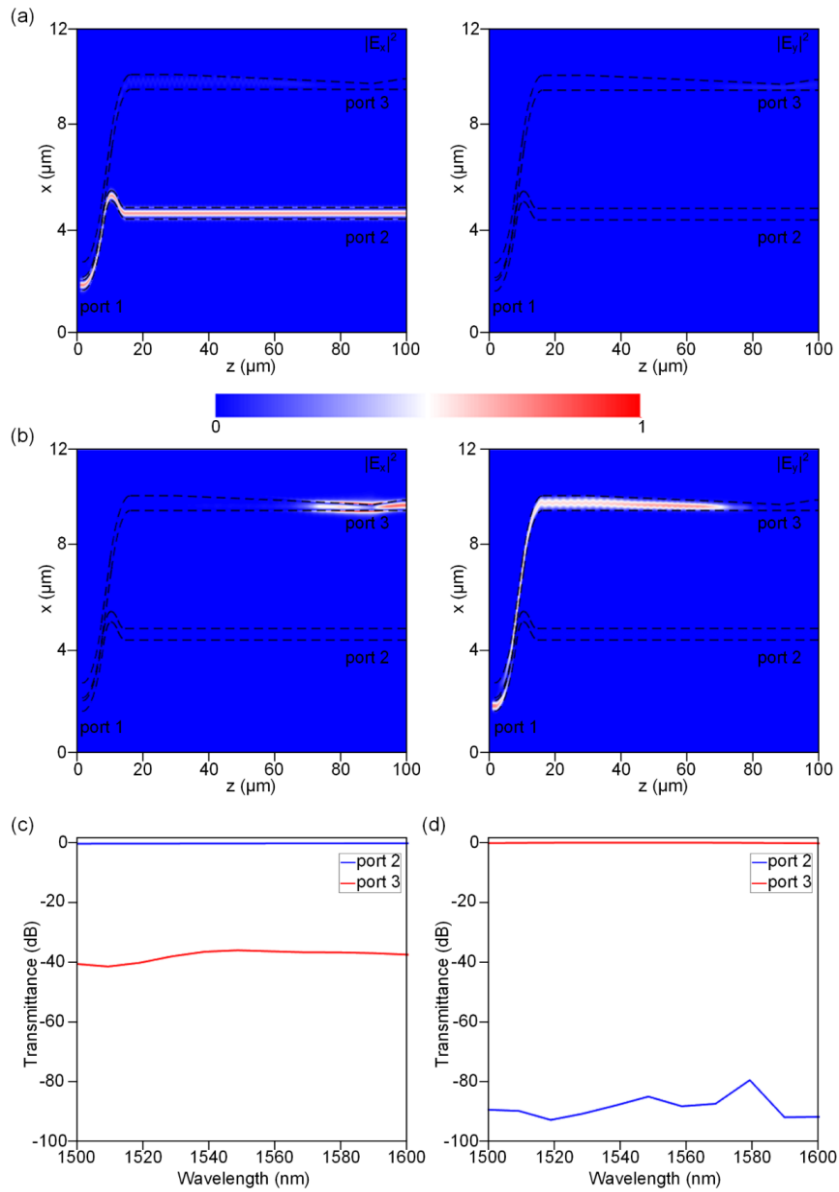

Supplementary Figure S8. Simulated results. (a, b) Field intensity distributions of $|E_x|^2$ and $|E_y|^2$ along the propagation direction, when (a) $TE_0$ and (b) $TM_0$ modes input from port 1. (c, d) Transmittance of the $TE_0$ mode at port 2 and port 3, when (c) $TE_0$ and (d) $TM_0$ modes input from port 1.

Supplementary Note 5: Simulated and experimental results with TE$_1$ and TM$_1$ input

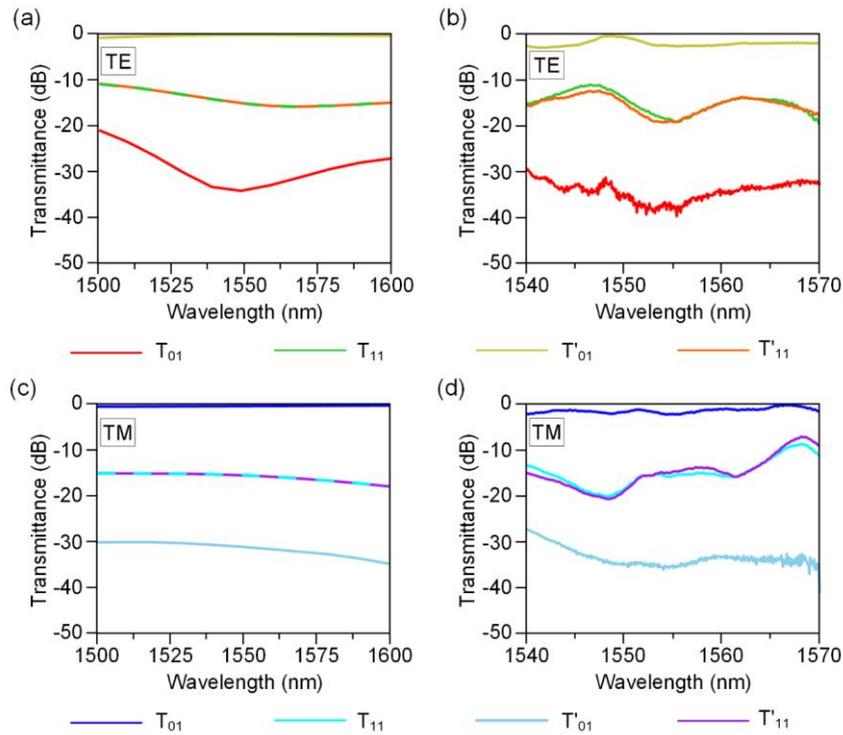

Supplementary Figure S9. Transmittance spectra. (a, c) Simulated transmittance spectra at the output port over the wavelength range of 1500–1600 nm. For TE$_0$ and TE$_1$ mode (a), TM$_0$ and TM$_1$ mode (c). (b) (d) Experimental transmittance spectra at the output port over the wavelength range of 1540–1570 nm. For TE$_0$ and TE$_1$ mode (b), TM$_0$ and TM$_1$ mode (d).

**Supplementary Note 5: Alignment error**

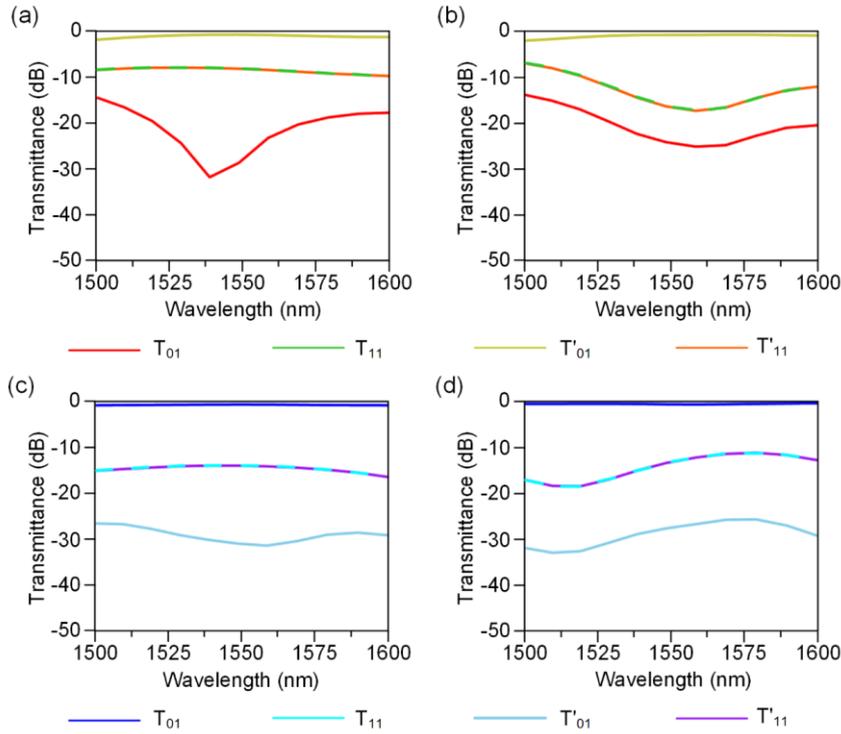

Supplementary Figure S10. Transmittance spectra. Simulated transmission spectra of the double-coupled silicon waveguides, when $TE_1$ (a, b) and $TM_1$ modes (c, d) are injected. (a, c) indicates a positive alignment error, i.e., $W_{Ra}$ is enhanced by 50 nm, and $W_{La}$ is decreased by 50 nm, with respect to the baseline geometrical parameters. (b, d) indicates a negative alignment error, i.e., $W_{Ra}$ is decreased by 50 nm, and $W_{La}$ is enhanced by 50 nm, with respect to the baseline geometrical parameters.

If the alignment position of the second-step EBL and the third-step EBL are not completely consistent, $W_{Ra}$ and $W_{La}$ may deviate from the predesigned values. Supplementary Fig. S10 demonstrates that the double-coupled silicon waveguides can maintain well transmission spectra in the wavelength range of interest, even if the alignment errors are within the range of ±50 nm.

## Supplementary Note 6: Shortened device

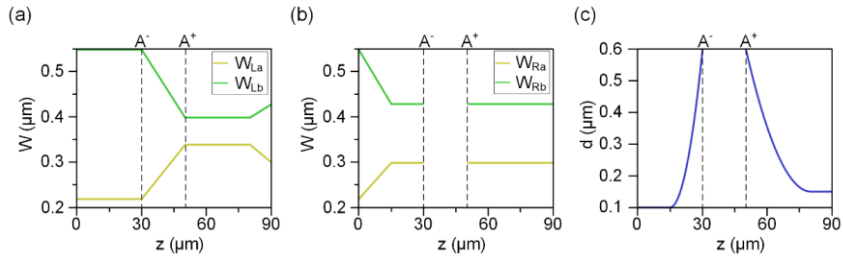

Supplemental Figure S11. Structural parameters of the shortened device. (a) Width of the left waveguide; (b) Width of the right waveguide; (c) Gap distance, versus z.

When alignment errors are not considered, simulations using the structural parameters shown in Supplementary Fig. S11 yield the transmission spectra presented in Supplementary Fig. S12. It can be observed that, after altering the structural parameters, the device length reduces from the 170nm length in the main text to 90 nm, while still achieving the expected results.

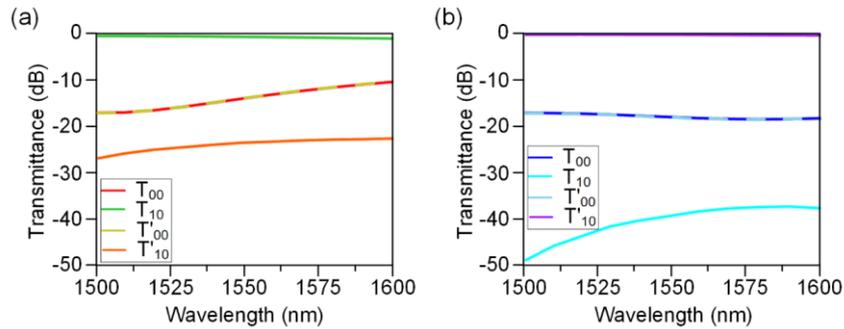

Supplementary Figure S12. Transmittance spectra. Simulated transmittance spectra for $TE_0$ (a) and $TM_0$ (b) mode is injected.